\documentclass[%
 reprint,
superscriptaddress,
 amsmath,amssymb,
]{revtex4-1}

\usepackage{ulem}
\usepackage{color}
\usepackage{graphicx}
\usepackage{dcolumn}
\usepackage{bm}

\begin{document}

\preprint{APS/123-QED}

\title{Variational cluster approach to $s$-wave pairing in heavy-fermion superconductors}

\author{Keisuke Masuda}
\email{masuda@kh.phys.waseda.ac.jp}
\affiliation{Department of Physics, Waseda University, Shinjuku, Tokyo 169-8555, Japan}
\author{Daisuke Yamamoto}
\affiliation{Waseda Institute for Advanced Study, Waseda University, Tokyo 169-8050, Japan}

\date{\today}

\begin{abstract}
We study $s$-wave Cooper pairing in heavy-fermion systems. We analyze the periodic Anderson model by means of the variational cluster approach (VCA) focusing on the interorbital Cooper pairing between a conduction electron ($c$ electron) and an $f$ electron, called the ``$c$-$f$ pairing.'' It is shown that the $s$-wave superconductivity appears coexisting with long-range antiferromagnetic order when electrons or holes are doped into the system at half filling. The antiferromagnetic order vanishes when the doping concentration exceeds a certain critical value, leading to a pure $s$-wave superconducting state. Moreover, the comparative study with different reference systems used in the VCA shows that the interorbital $c$-$f$ pairing is essential for the appearance of the $s$-wave superconductivity.

\begin{description}
\item[PACS numbers]
74.70.Tx, 74.20.Mn, 74.25.Dw
\end{description}
\end{abstract}

\pacs{Valid PACS appear here}
\maketitle


\section{\label{sec1}introduction}
Heavy-fermion systems have provided opportunities to study various types of superconductivity. For example, the Ce-based compound ${\rm CeCoIn_{5}}$ has two kinds of superconducting states, one of which observed in higher magnetic fields is a strong candidate for the Fulde-Ferrell-Larkin-Ovchinnikov state with finite center-of-mass momentum of the Cooper pairs \cite{Matsuda_Shimahara}. In superconductors without inversion symmetry, such as ${\rm CePt_{3}Si}$ and ${\rm CeRhSi_{3}}$, the exotic parity mixing between spin-singlet and spin-triplet states is expected to occur due to the existence of the antisymmetric spin-orbit interaction \cite{Bauer,Bauer_Kaldarar,Pfleiderer}. The coexistence of superconductivity and long-range magnetic order has been observed in several ferromagnets (${\rm UGe_{2}}$, URhGe, etc.) as well as in several antiferromegnets (${\rm UPd_{2}Al_{3}}$, ${\rm UNi_{2}Al_{3}}$, etc.) \cite{Pfleiderer}. A variety of experimental and theoretical efforts have been devoted to understanding those exotic states.

Superconductivity with simple $s$-wave pairing symmetry is another intriguing phenomenon in heavy-fermion systems. Usually, heavy-fermion superconductors favor the nodal pairing states, such as the $d$-wave and $p$-wave states, rather than the $s$-wave state. This is because the strong Coulomb repulsion in those systems is incompatible with intrasite Cooper pairing, which gives the nodal $d$-wave and $p$-wave states. In fact, nuclear resonance [NMR and nuclear quadrupole resonance (NQR)] experiments have demonstrated that many of the heavy-fermion superconductors possess the nodal superconducting gaps \cite{Pfleiderer,Mito,Kohori}. On the other hand, some heavy-fermion compounds, such as ${\rm CeRu_{2}}$ \cite{Matsuda_Kohori,Mukuda,Kiss,Sakakibara_Yamada,Kittaka_Sakakibara}, ${\rm CeCo_{2}}$ \cite{Ishida_Mukuda}, and the recently reinvestigated ${\rm CeCu_{2}Si_{2}}$ \cite{Kittaka_Aoki}, are known to exhibit $s$-wave superconductivity. In the BCS theory, such $s$-wave superconductivity is explained as a result of the electron-phonon attraction between electrons. However, as mentioned above, heavy-fermion compounds have the strong Coulomb repulsion, which is considered as the dominant interaction between electrons. Thus, the $s$-wave superconductivity in those compounds may come from another mechanism.

The multiorbital nature is one of the characteristic features of heavy-fermion systems, which are composed of itinerant electrons in the conduction orbitals ($c$ electrons) and localized electrons in the $f$ orbitals ($f$ electrons). The correlation between $c$ and $f$ electrons leads to various intriguing phenomena, such as the Kondo effect \cite{Pfleiderer,Hewson}, quantum critical behavior \cite{Stewart_hf,Nakatsuji,Shishido}, and magnetic orderings due to the Ruderman-Kittel-Kasuya-Yosida interaction \cite{Hewson,Amato}. Recently, the importance of such orbital degrees of freedom has also been recognized in the studies of superconductivity in the other strongly correlated electron systems. For example, the material dependence in the critical temperature of cuprates has been explained by using the multiorbital Hubbard models \cite{Sakakibara_Usui}. Moreover, the multiorbital nature is considered to be the key for understanding the high-$T_{\rm c}$ superconducting properties in iron pnictides \cite{Stewart_iron}. Previous studies \cite{Hanzawa_Yosida,Spalek,Masuda_Yamamoto} suggested that the multiorbital nature can be a source of $s$-wave superconductivity in heavy-fermion systems. Hanzawa and Yosida \cite{Hanzawa_Yosida} and Spa{\l}ek \cite{Spalek} discussed the interorbital Cooper pairing between $c$ and $f$ electrons, which we call the ``$c$-$f$ pairing,'' as a possible mechanism for $s$-wave superconductivity. They estimated the order of the critical temperature in the periodic Anderson model with infinitely large Coulomb repulsion. More recently, the present authors \cite{Masuda_Yamamoto} also studied the $c$-$f$ pairing for finite Coulomb repulsion, and presented a mean-field phase diagram of the $s$-wave superconducting state. Note, however, that the mean-field approximation cannot properly describe local charge, spin, and orbital fluctuation effects, which are crucial in heavy-fermion systems. Thus more sophisticated treatment is required to achieve a deeper understanding of the nature of the interorbital pairing.

In this paper, we use the variational cluster approach (VCA) \cite{Potthoff_Aichhorn} to study $s$-wave superconductivity in heavy-fermion systems. The VCA can properly take into account the local Coulomb repulsion \cite{Balzer_EPL,Balzer_PRB} and allows us to deal with various long-range orders, such as charge-density-wave \cite{Aichhorn_Evertz}, $d$-wave superconducting \cite{Senechal_PRL,Sahebsara}, and antiferromagnetic \cite{Dahnken_Aichhorn,Horiuchi} orders. Here, we apply the VCA to the standard periodic Anderson model considering all three types of $s$-wave Cooper pairings, i.e., between $c$ electrons ($c$-$c$ pairing), between $f$ electrons ($f$-$f$ pairing), and between $c$ and $f$ electrons ($c$-$f$ pairing). We also consider possible antiferromagnetic order, which has been shown to emerge when the Coulomb repulsion is sufficiently strong \cite{Horiuchi,Rozenberg,Vekic}. We calculate those order parameters and find five different phases depending on the parameters. At half filling, the system undergoes a second-order phase transition from nonmagnetic Kondo insulator to antiferromagnetic state when we increase the Coulomb repulsion. Away from half filling, we find the $s$-wave superconducting phase, in which all the superconducting order parameters ($c$-$c$, $f$-$f$, and $c$-$f$ pairings) have finite values. We also find the coexistence phase of the $s$-wave superconductivity and long-range antiferromagnetic order in a region closer to half filling. In the VCA, the self-energy of the original system is approximated by that of a reference system consisting of isolated clusters. An advantage of the VCA is that it can treat symmetry-breaking states by assuming effective fields called the ``Weiss fields'' in the reference system. We compare two different reference systems with and without the Weiss field that acts as a pair potential for the $c$-$f$ pairing, and conclude that the formation of Cooper pairs between $c$ and $f$ electrons is indeed an essential mechanism to stabilize the $s$-wave superconducting states in heavy-fermion systems.

The remainder of the paper is organized as follows. In Sec. \ref{sec2}, we introduce the periodic Anderson model and extend the formulation of the VCA to describe the $s$-wave superconductivity in the model. In Sec. \ref{sec3}, we show the phase diagram obtained by the VCA. In Sec. \ref{sec4}, the mechanism for the emergence of the $s$-wave superconducting states is discussed. The final section, Sec. \ref{sec5}, is devoted to conclusions.

\section{\label{sec2}model and method}
We consider the periodic Anderson model, which is believed to capture the essential physics of heavy-fermion systems. The Hamiltonian of the model is given by
\begin{eqnarray}
\nonumber H_{\rm PAM}&=&-t\sum_{\langle{ij}\rangle}\sum_{\sigma}(c^{\dagger}_{i\sigma}c_{j\sigma}+{\rm H.c.})+\epsilon_{f}\sum_{i\sigma}n^{f}_{i\sigma}\\
\nonumber &&-V\sum_{i\sigma}(f^{\dagger}_{i\sigma}c_{i\sigma}+{\rm H.c.})+U\sum_{i}n^{f}_{i\uparrow}n^{f}_{i\downarrow}\\
&&-\mu\sum_{i\sigma}(n^{c}_{i\sigma}+n^{f}_{i\sigma}),\label{H_original}
\end{eqnarray}
where $c^{\dagger}_{i\sigma}$ ($f^{\dagger}_{i\sigma}$) creates an itinerant $c$ electron (a localized $f$ electron) with spin $\sigma$ at site $i$, $n^{c}_{i\sigma}=c^{\dagger}_{i\sigma}c_{i\sigma}$, and $n^{f}_{i\sigma}=f^{\dagger}_{i\sigma}f_{i\sigma}$. Here, $t$ is the hopping amplitude of $c$ electrons, $\epsilon_{f}$ is the on-site energy of $f$ electrons, $V$ is the hybridization between $c$ and $f$ states, $U$ is the on-site Coulomb repulsion in the $f$ orbital, and $\mu$ is the chemical potential. The sum $\langle{ij}\rangle$ is taken over nearest-neighbor pairs of lattice sites. We consider the system on a square lattice in this study.

We study the model (\ref{H_original}) using the VCA \cite{Potthoff_Aichhorn}, which is based on the self-energy functional theory (SFT) proposed by Potthoff \cite{Potthoff_EPJB}. We first assume a reference system that is given as a set of identical clusters $\Gamma$ of two neighboring sites. The Hamiltonian of the reference system is $H'=\sum_{\Gamma}H'_{\Gamma}$,
\begin{equation}
H'_{\Gamma}=H'_{\rm PAM}+H'_{cc}+H'_{ff}+H'_{cf}+H'_{\rm AF},\label{H_cluster}
\end{equation}
where
\begin{eqnarray}
\nonumber H'_{\rm PAM}&=&-t\sum_{\langle{ij}\rangle \in \Gamma, \sigma }(c^{\dagger}_{i\sigma}c_{j\sigma}+{\rm H.c.})+\epsilon_{f}\sum_{i \in \Gamma, \sigma}n^{f}_{i\sigma}\\
\nonumber &&-V\sum_{i \in \Gamma, \sigma}(f^{\dagger}_{i\sigma}c_{i\sigma}+{\rm H.c.})+U\sum_{i \in \Gamma}n^{f}_{i\uparrow}n^{f}_{i\downarrow}\\
&&-\mu'\sum_{i \in \Gamma, \sigma}(n^{c}_{i\sigma}+n^{f}_{i\sigma}),\\
H'_{cc}&=&h'_{cc}\sum_{i \in \Gamma}(c_{i \uparrow}c_{i \downarrow}+{\rm H.c.}),\\
H'_{ff}&=&-h'_{ff}\sum_{i \in \Gamma}(f_{i \uparrow}f_{i \downarrow}+{\rm H.c.}),\\
H'_{cf}&=&-h'_{cf}\sum_{i \in \Gamma}(c_{i \uparrow}f_{i \downarrow}-c_{i \downarrow}f_{i \uparrow}+{\rm H.c.}),\\
H'_{\rm AF}&=&-h'_{\rm AF}\sum_{i \in \Gamma} e^{i {\bf Q} \cdot {{\bf r}_{i}}} (n^{f}_{i\uparrow}-n^{f}_{i\downarrow}).
\end{eqnarray}
Here, $\Gamma$ is the label of each cluster and ${\bf Q}$ is the commensurate wave vector $(\pi,\pi)$. As shown in Eq. (\ref{H_cluster}), the cluster Hamiltonian $H'_{\Gamma}$ includes four types of Weiss-field terms, $H'_{cc}$, $H'_{ff}$, $H'_{cf}$, and $H'_{\rm AF}$. The first three terms allow for describing the $c$-$c$, $f$-$f$, and $c$-$f$ pairing orders, respectively. The last term gives long-range antiferromagnetic order. The corresponding Weiss fields, $h'_{cc}$, $h'_{ff}$, $h'_{cf}$, and $h'_{\rm AF}$, are determined by the variational conditions as mentioned below. To keep the thermodynamic consistency \cite{Aichhorn_Arrigoni_2,Senechal_arXiv}, the cluster chemical potential $\mu'$ is also treated as a variational parameter. We denote the set of these variational parameters as ${\bf t'}\equiv(h'_{cc}, h'_{ff}, h'_{cf}, h'_{\rm AF}, \mu')$. We assume that the Weiss field $h'_{\rm AF}$ acts only on $f$ electrons, which is justified by the fact that the antiferromagnetic order in this system is mainly due to the Coulomb repulsion between $f$ electrons.

We introduce the following Nambu spinor defined on each cluster:
\begin{equation}
{\bm \Psi}\!\!=\!(c_{1\uparrow}, c_{2\uparrow}, f_{1\uparrow}, f_{2\uparrow},c^{\dagger}_{1\downarrow}, c^{\dagger}_{2\downarrow}, f^{\dagger}_{1\downarrow}, f^{\dagger}_{2\downarrow})^{\rm T},
\end{equation}
where the two sites on the cluster $\Gamma$ are labeled $1$ and $2$. By diagonalizing the two-site Hamiltonian $H'_{\Gamma}$, we can easily obtain the Green's-function matrix ${\bf G}'=\langle\langle {\bm \Psi};{\bm \Psi}^{\dagger} \rangle\rangle$ and the grand potential $\Omega'$ of the reference system $H'$. Note that ${\bf G}'$ includes the anomalous Green's functions regarding the $c$-$c$, $f$-$f$, and $c$-$f$ pairings as the off-diagonal components. We can also calculate the self-energy matrix ${\bf \Sigma}'$ of the reference system by using ${\bf \Sigma}'({\bf t'})={{\bf G}'_{0}}^{-1}-{{\bf G}'}^{-1}$, where ${\bf G}'_{0}$ is the free Green's function of the reference system obtained by setting $U=0$ in Eq. (\ref{H_cluster}).

According to the SFT \cite{Potthoff_EPJB}, the grand potential of the original system can be written as
\begin{eqnarray}
\nonumber \Omega\,({\bf t'}) &=& \Omega' -\frac{N}{2}\, {\rm Tr} \ln{[-{\bf G'}]}\\
&+& \sum_{\tilde {\bf k}} {\rm Tr} \ln{[ -{\bf G}_{\rm VCA}({\tilde {\bf k}}) ]}\,-2N(\mu-\mu'),\label{gpotential}
\end{eqnarray}
where $N$ is the total number of lattice sites. In the VCA \cite{Potthoff_Aichhorn}, the self-energy of the original system is approximated by that of the reference system as ${\bf G}_{\rm VCA}({\tilde {\bf k}})\equiv( {\bf G}_{0}({\tilde {\bf k}})^{-1}-{\bm \Sigma}' )^{-1}$. Here, ${\bf G}_{0}({\tilde {\bf k}})$ is the free Green's function of the original system (\ref{H_original}) with ${\tilde {\bf k}}$ being the wave vector in the Brillouin zone of the reference system. The last term on the right-hand side of Eq. (\ref{gpotential}), $-2N(\mu-\mu')$, arises from the anticommutation relation of electron operators when we rewrite the Hamiltonians, Eqs. (\ref{H_original}) and (\ref{H_cluster}), using the Nambu spinor. Practical details of the evaluation of Eq. (\ref{gpotential}) are given in the Appendix. We determine the optimal values of the variational parameters ${\bf t'}_{\rm opt}$ by solving the variational problems $\frac{\partial \Omega}{\partial h'_{cc}}=0$, $\frac{\partial \Omega}{\partial h'_{ff}}=0$, $\frac{\partial \Omega}{\partial h'_{cf}}=0$, $\frac{\partial \Omega}{\partial h'_{\rm AF}}=0$, and $\frac{\partial \Omega}{\partial \mu'}=0$, simultaneously. For a given total density $n$, we also determine the chemical potential $\mu$ so that it can satisfy the number equation $n-\sum_{i\sigma} \langle n^{c}_{i\sigma}+n^{f}_{i\sigma} \rangle/N =0$, where the average $\langle{\cdots}\rangle$ is calculated from the VCA Green's function with the optimized variational parameters ${\bf G}_{\rm VCA}({\tilde {\bf k}})|_{{\bf t'}={\bf t'}_{\rm opt}}$. The condition $\frac{\partial \Omega}{\partial \mu'}=0$ guarantees that the thermodynamic relation $n=-\frac{1}{N}\frac{\partial \Omega}{\partial \mu}$ is satisfied \cite{Aichhorn_Arrigoni_2,Senechal_arXiv}. Using the same Green's function ${\bf G}_{\rm VCA}({\tilde {\bf k}})|_{{\bf t'}={\bf t'}_{\rm opt}}$, we evaluate the following quantities:
\begin{eqnarray}
\Delta_{cc}&=&\frac{1}{N}\sum_{i} \langle c_{i\downarrow}c_{i\uparrow} \rangle ,\\
\Delta_{ff}&=&\frac{1}{N}\sum_{i} \langle f_{i\downarrow}f_{i\uparrow} \rangle ,\\
\Delta_{cf}&=&\frac{1}{2N}\sum_{i} \langle c_{i\downarrow}f_{i\uparrow}-c_{i\uparrow}f_{i\downarrow} \rangle ,\\
m_{c}&=& \frac{1}{2N}\sum_{i} e^{i {\bf Q} \cdot {{\bf r}_{i}}} \langle n^{c}_{i\uparrow}-n^{c}_{i\downarrow} \rangle  ,\\
m_{f}&=&\frac{1}{2N}\sum_{i} e^{i {\bf Q} \cdot {{\bf r}_{i}}} \langle n^{f}_{i\uparrow}-n^{f}_{i\downarrow} \rangle ,\\
\delta_{cf}&=&\frac{1}{2N}\sum_{i} e^{i {\bf Q} \cdot {{\bf r}_{i}}} \langle c_{i\downarrow}f_{i\uparrow}+c_{i\uparrow}f_{i\downarrow} \rangle ,\label{delta_cf}
\end{eqnarray}
where $\Delta_{cc}$, $\Delta_{ff}$, and $\Delta_{cf}$ represent the $s$-wave superconducting order parameters for the $c$-$c$, $f$-$f$, and $c$-$f$ pairings, respectively, and $m_{c}$ ($m_{f}$) is the staggered magnetization in the $c$ ($f$) orbital. The quantity $\delta_{cf}$ represents a staggered modulation of the difference between the anomalous average $\langle c_{i\downarrow}f_{i\uparrow} \rangle$ and its time-reversal counterpart $-\langle c_{i\uparrow}f_{i\downarrow} \rangle$. Throughout this work, we fix the value of $\epsilon_{f}$ to $-U/2$, considering the situation where the Fermi level is located near the center of the upper and lower Hubbard bands of $f$ electrons. Under the symmetric condition $\epsilon_{f}=-U/2$, the models for electron-doped ($n=2.0+\delta$) and hole-doped ($n=2.0-\delta$) systems are symmetric with each other about half filling ($n=2.0$). Thus, we discuss only the electron-doped case hereafter. We set the hybridization $V=t$ and the temperature $T=0$ in the present study.

\section{\label{sec3}results}
\begin{figure}[b]
\includegraphics[width=7.8cm]{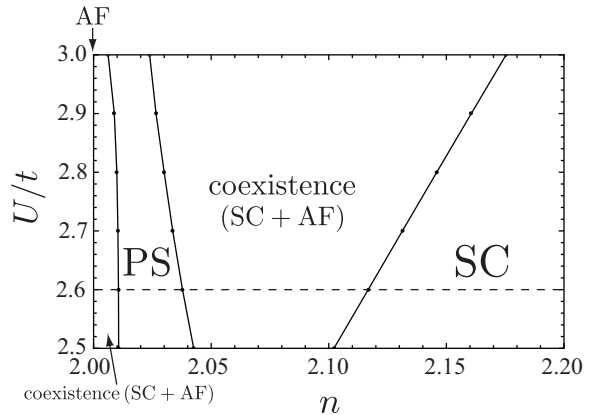}
\caption{\label{phase_diagram} The phase diagrams in the $(n,U/t)$ plane. The horizontal dashed line indicates the line of $U/t=2.6$.}
\end{figure}
\begin{figure}[t]
\includegraphics[width=7.0cm]{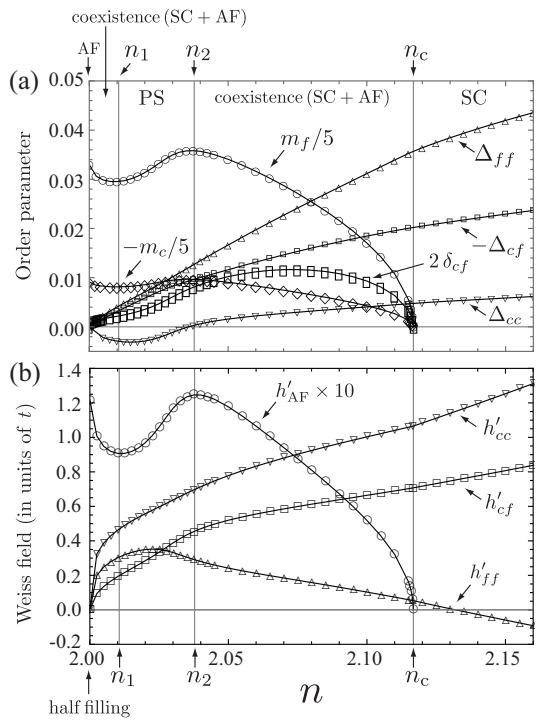}
\caption{\label{OP-WF} The $n$ dependencies of (a) the order parameters and (b) the Weiss fields for $U/t=2.6$.}
\end{figure}

At half filling, the system exhibits the Kondo insulating state, which changes into an antiferromagnetic state when the Coulomb repulsion exceeds a critical value $U_{\rm c}/t$ \cite{Horiuchi,Rozenberg,Vekic}. Our VCA analysis gives $U_{\rm c}/t\,{\approx}\,2.31$. In the following, we focus on the case of $U>U_{\rm c}$. Figure \ref{phase_diagram} shows the phase diagram in the ($n$,\,$U/t$) plane. To explain each phase in the phase diagram, we show in Fig. \ref{OP-WF}(a) the $n$ dependencies of the order parameters at $U/t=2.6$, which is marked by the horizontal dashed line in Fig. \ref{phase_diagram}. We also show the corresponding behavior of the Weiss fields in Fig. \ref{OP-WF}(b). Away from half filling, only the superconducting order parameters, $\Delta_{ff}$, $\Delta_{cf}$, and $\Delta_{cc}$, have finite values, which means that the system is in the pure $s$-wave superconducting (SC) phase. The values of the order parameters satisfy the inequality $\Delta_{ff}>\Delta_{cf}>\Delta_{cc}$.
\begin{figure}[t]
\includegraphics[width=7.4cm]{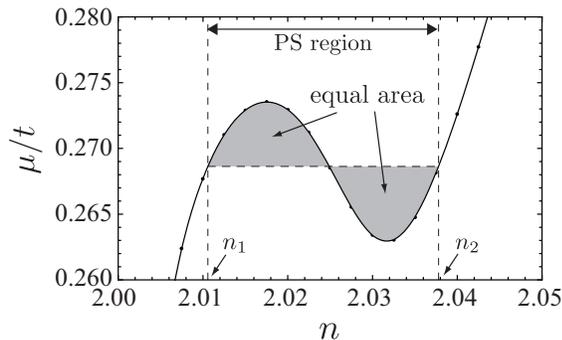}
\caption{\label{Maxwell-cr} The $n$ dependence of the chemical potential for $U/t=2.6$.}
\end{figure}
When we decrease the density $n$, the staggered magnetizations $m_{f}$ and $m_{c}$ appear at the critical density $n_{\rm c}\,{\approx}\,2.12$, below which the $s$-wave superconductivity coexists with the antiferromagnetic (AF) order. The quantity $\delta_{cf}$ takes a nonzero value only when the antiferromagnetic order occurs ($m_{c},m_{f} \ne 0$), as will be explained later. When $n$ is decreased further, the system exhibits phase separation (PS). Since the difference of the grand potentials at $\mu=\mu_{\rm A}$ and $\mu=\mu_{\rm B}$ was given by $\Delta\Omega=-N\int^{\mu_{\rm B}}_{\mu_{\rm A}} n(\mu)d\mu$, we determined the boundaries $n_{1}$ and $n_{2}$ of the PS region from the Maxwell construction in the $(n,\,\mu/t)$ plane, as shown in Fig. \ref{Maxwell-cr}. This type of phase separation was also found in the previous VCA studies that discussed the coexistence of $d$-wave superconductivity and antiferromagnetic order in the Hubbard model \cite{Aichhorn_Arrigoni_1,Aichhorn_Arrigoni_2,Aichhorn_Arrigoni_3,Aichhorn_Arrigoni_4}. One of these studies \cite{Aichhorn_Arrigoni_1} has predicted that the PS region becomes narrower as the cluster size increases and may vanish in the limit of large cluster size. This may also be the case for the PS region in our results. Finally, near half filling ($2\,{\leq}\,n\!<\!n_{1}$), the system exhibits the coexistence phase again.

\begin{figure}[t]
\includegraphics[width=7.8cm]{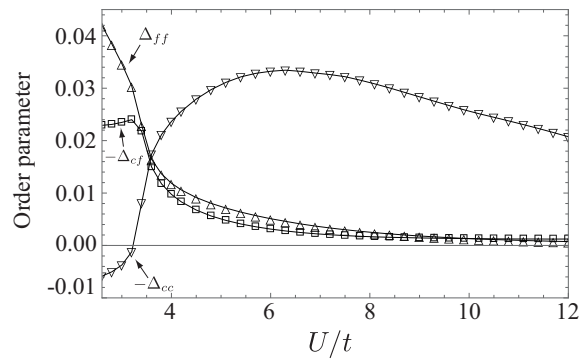}
\caption{\label{Udep_n215} The $U$ dependencies of the superconducting order parameters for $n=2.15$.}
\end{figure}

For $U/t=2.6$, the $f$-$f$ pairing amplitude $|\Delta_{ff}|$ is larger than the other ones, $|\Delta_{cc}|$ and $|\Delta_{cf}|$, as shown in Fig. \ref{OP-WF}(a). This is attributed to the large density of states (DOS) for $f$ electrons \cite{Mutou}. In usual heavy-fermion compounds, the Coulomb repulsion $U$ ($\sim 5$--$12$\,\,${\rm eV}$ \cite{Hewson}) is quite large compared to the hopping $t$ and the hybridization $V$. In such a situation ($U\! \gg \! t,V$), the on-site $f$-$f$ pairing is expected to be strongly suppressed. Figure \ref{Udep_n215} shows the $U$ dependencies of each superconducting order parameter in the superconducting state. The magnitude relation among $\Delta_{cc}$, $\Delta_{cf}$, and $\Delta_{ff}$ drastically changes at $U/t \sim 3.5$, and $|\Delta_{cc}|$ becomes much larger than the others in a very large $U$ region. Therefore, the $c$-$c$ pairing is dominant in a realistic parameter regime. However, the values of $\Delta_{cf}$ and $\Delta_{ff}$ do not become completely zero due to the hybridization $V$, and especially the formation of the $c$-$f$ pairing is essentially important for the $s$-wave superconducting state as will be explained in the next section.

\section{\label{sec4}discussion}
We investigate here the role of the $c$-$f$ pairing in the formation of $s$-wave superconducting state by a comparative study with the VCA. To this end, we carry out additional calculations based on the following cluster Hamiltonians instead of Eq. (\ref{H_cluster}): (i) $H'_{\Gamma,1}=H'_{\rm PAM}+H'_{cc}+H'_{ff}$ (i.e., the $c$-$f$ pairing field is not considered); (ii) $H'_{\Gamma,2}=H'_{\rm PAM}+H'_{cf}$ (i.e., only the $c$-$f$ pairing field is considered). 

In the first case (i), we found only a trivial solution $h'_{cc}=h'_{ff}=0$, namely, no superconducting solution is obtained ($\Delta_{cc}=\Delta_{ff}=\Delta_{cf}=0$). This indicates that the occurrence of the $s$-wave superconductivity requires the $c$-$f$ pairing field $h'_{cf}$, i.e., the $c$-$f$ pairing plays a crucial role in the mechanism for the $s$-wave superconductivity. Indeed, in the second case (ii), we find a solution with $h'_{cf}\ne0$ and $\Delta_{cf}\ne0$. Note that the other order parameters $\Delta_{cc}$ and $\Delta_{ff}$ also have finite values even though the corresponding Weiss fields $h'_{cc}$ and $h'_{ff}$ are not taken into account. This stems from the hybridization $V$ between $c$ and $f$ states. Due to the existence of $h'_{cf}$ and $V$, the self-energy ${\bf \Sigma}'({\bf t'})$ has the off-diagonal components $\Sigma'_{cc}({\bf t'})$ for the $c$-$c$ pairing and $\Sigma'_{ff}({\bf t'})$ for the $f$-$f$ pairing as well as $\Sigma'_{cf}({\bf t'})$ for the $c$-$f$ pairing, through the diagonalization of $H'_{\Gamma,2}$. Thus, all the superconducting order parameters, $\Delta_{cc}$, $\Delta_{ff}$, and $\Delta_{cf}$, have finite values although $H'_{\Gamma,2}$ does not include $h'_{cc}$ and $h'_{ff}$. This comparative study indicates that the pair potential for the $c$-$f$ pairing is essential for the occurrence of the $s$-wave superconductivity.

Let us discuss the mechanism giving rise to the effective $c$-$f$ pair potential in the periodic Anderson model (\ref{H_original}). When the Coulomb repulsion $U$ is quite strong, the physics of the system may be understood in a perturbative fashion \cite{Masuda_Yamamoto}. Assuming that the repulsion $U$ is much larger than the hybridization $V$, we derived an effective Hamiltonian of the periodic Anderson model through the Schrieffer-Wolff transformation in the previous work \cite{Masuda_Yamamoto}. The effective Hamiltonian includes the direct and spin-exchange interactions between $c$ and $f$ electrons. The first one describes the charge fluctuation in the $c$ orbital depending on the occupation state in the $f$ orbital and the second one represents the spin fluctuation between $c$ and $f$ orbitals. We concluded in Ref. \cite{Masuda_Yamamoto} that these interorbital perturbative processes play the role of a glue for the $c$-$f$ pairing.

\begin{figure}[t]
\includegraphics[width=6.9cm]{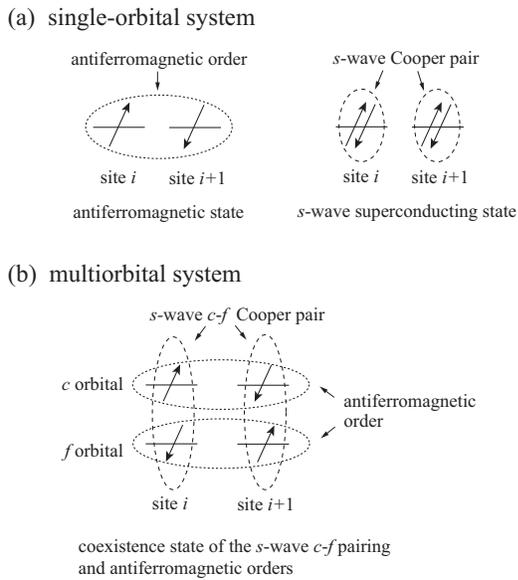}
\caption{\label{schematic_fig} Schematic pictures on the relationship between $s$-wave superconductivity and antiferromagnetism. Panels (a) and (b) correspond to the cases of single-orbital and multiorbital systems, respectively.}
\end{figure}
Finally, we consider the reason why the $s$-wave superconductivity can coexist with the long-range antiferromagnetic order near the half filling (see Fig. \ref{phase_diagram}). In usual single-orbital systems, on-site Cooper pairing and antiferromagnetism compete with each other since the local spin polarization is incompatible with the formation of local spin singlets [see Fig. \ref{schematic_fig}(a)]. However, the interorbital pairing in the present case does not suffer from such incompatibility. As seen in Fig. \ref{schematic_fig}(b), antiferromagnetic order occurs in each of the $c$ and $f$ orbitals, between which on-site $s$-wave Cooper pairs can be formed. Note that since the antiferromagnetic order breaks the local spin up-down symmetry, the anomalous average $\langle c_{i\downarrow}f_{i\uparrow} \rangle$ and its time-reversal counterpart $-\langle c_{i\uparrow}f_{i\downarrow} \rangle$ have different values. For example, in Fig. \ref{schematic_fig}(b), $|\langle c_{i\uparrow}f_{i\downarrow} \rangle|>|\langle c_{i\downarrow}f_{i\uparrow} \rangle|$ for site $i$ and $|\langle c_{i+1\downarrow}f_{i+1\uparrow} \rangle|>|\langle c_{i+1\uparrow}f_{i+1\downarrow} \rangle|$ for site $i\!+\!1$. Therefore, the difference $\delta_{cf}$ defined in Eq. (\ref{delta_cf}) has a finite value in the coexistence phase of the $c$-$f$ pairing and antiferromagnetic orders.

\section{\label{sec5}conclusion}
We have investigated $s$-wave superconductivity in heavy-fermion systems in terms of the variational cluster approach (VCA) to the periodic Anderson model. In the VCA, we have taken into account all the three types of $s$-wave Cooper pairings: the intraorbital pairings between $c$ electrons and between $f$ electrons, and the interorbital pairing between $c$ and $f$ electrons. We have shown that $s$-wave superconducting states appear when electrons or holes are doped to the system at half filling. In a region close to half filling, the $s$-wave superconductivity coexists with long-range antiferromagnetic order. The VCA comparative analysis with different reference systems indicated that the $c$-$f$ pairing plays a dominant role in the formation of the $s$-wave superconducting state. These results might advance the understanding of the fully gapped superconducting states observed in several Ce-based materials \cite{Matsuda_Kohori,Mukuda,Kiss,Sakakibara_Yamada,Kittaka_Sakakibara,Ishida_Mukuda,Kittaka_Aoki}.

Recently, $s$-wave superconductivity in heavy-fermion systems has also been studied with the Kondo-lattice model \cite{Bodensiek}, in which $f$ electrons are assumed to be almost localized and have only spin degrees of freedom. The authors of Ref. \cite{Bodensiek} have shown that the correlation between the localized spins and conduction electrons through the Kondo exchange coupling gives rise to local pairing interaction, leading to $s$-wave superconductivity. It is known that the periodic Anderson model studied in the present work is mapped onto the Kondo lattice model in the so-called Kondo limit \cite{Schrieffer,Tsunetsugu,Sinjukow}. The relation between the $s$-wave superconducting states proposed in the two models remains an intriguing issue for future work.

\begin{acknowledgments}
We would like to thank T. Shirakawa, H. Watanabe, and S. Yunoki for useful discussions. The calculations were performed by using the RIKEN Integrated Cluster of Clusters (RICC) facility. One of the authors (K.M.) was supported by a Grant-in-Aid from JSPS and by a Grant for Excellent Graduate Schools, MEXT, Japan. This work was also supported by KAKENHI grants from JSPS (Grant No. 26800200) (D.Y.).
\end{acknowledgments}

\appendix

\section{\label{ape}Evaluation of the grand potential}
We evaluate the grand potential given by Eq. (\ref{gpotential}) at $T=0$. We first introduce the matrices ${\bf Q}^{(e)}$ and ${\bf Q}^{(h)}$ whose elements are given by \cite{Senechal_arXiv}
\begin{equation}
Q^{(e)}_{m,n}=\langle0|\Psi_{m}|n\rangle,\,\,\,
Q^{(h)}_{m,n}=\langle n|\Psi_{m}|0\rangle,\label{defQ}
\end{equation}
with
\begin{equation}
H'_{\Gamma}|0\rangle=E_{0}|0\rangle,\,\,\,
H'_{\Gamma}|n\rangle=E_{n}|n\rangle.
\end{equation}
Here, $|0\rangle$ ($|n\rangle$) is the ground ($n$th excited) state of the cluster Hamiltonian $H'_{\Gamma}$ and $\Psi_{m}$ is the $m$th component of the Nambu spinor ${\bm \Psi}$. Note that the excited states with even (odd) numbers of electrons can be ignored when the ground state $|0\rangle$ consists of even (odd) numbers of electrons. Thus, the number of excited states that have to be considered is $N_{e}=4^4/2$ and the size of the matrices ${\bf Q}^{(e)}$ and ${\bf Q}^{(h)}$ is $8{\times}N_{e}$ in the present two-site reference system with two orbitals per site. Using ${\bf Q}^{(e)}$ and ${\bf Q}^{(h)}$, we define the $8\,\times\,2N_e$ $Q$-matrix ${\bf Q}$ which has the following elements:
\begin{eqnarray}
Q_{m,l}=\left\{ \begin{array}{ll}
Q^{(e)}_{m,l} & (1 \leq l \leq N_e) \\
Q^{(h)}_{m,l-N_{e}} & (N_e+1 \leq l \leq 2N_e).\\
\end{array} \right.
\end{eqnarray} 
We also introduce the $2N_e\times2N_e$ diagonal matrix ${\bm \Lambda}$ whose diagonal elements are given by
\begin{eqnarray}
\Lambda_{l,l}=\left\{ \begin{array}{ll}
E_{l}-E_{0} & (1 \leq l \leq N_e) \\
-E_{l-N_e}+E_{0} & (N_e+1 \leq l \leq 2N_e).\\
\end{array} \right.
\end{eqnarray} 
In the Lehmann representation, the cluster Green's function ${\bf G'}(\omega)$ can be written as \cite{Aichhorn_Arrigoni_1,Senechal_arXiv}
\begin{equation}
{\bf G'}(\omega)={\bf Q}\,{\bf g}(\omega)\,{\bf Q}^{\dagger},\label{Gcl_1}
\end{equation}
where ${\bf g}(\omega)=(\omega-{\bm \Lambda})^{-1}$.

Note that the Tr in Eq. (\ref{gpotential}) includes the summation over the fermionic Matsubara frequencies \cite{Potthoff_EPJB}. We can rewrite the second term on the right-hand side of Eq. (\ref{gpotential}) as follows \cite{Aichhorn_Arrigoni_1,Senechal_arXiv}:
\begin{equation}
-\frac{N}{2}\,{\rm Tr} \ln{[ -{\bf G'} ]}=-\frac{N}{2}\,\sum^{2N_e}_{l=1}\omega'_{l}\Theta(-\omega'_{l})+R\label{Gcl_2},
\end{equation}
where $\omega'_{l}$ is the pole of the cluster Green's function (\ref{Gcl_1}), and $\Theta(x)$ is Heaviside step function defined by $\Theta(x)=1$ for $x\geq0$ and $\Theta(x)=0$ for $x<0$. The last term $R$ represents the contribution from the poles of the self-energy ${\bm \Sigma}'$. We note that since $\omega'_{l}$ is given by the diagonal elements of ${\bf \Lambda}$, the first term of Eq. (\ref{Gcl_2}) is simplified as
\begin{equation}
-\frac{N}{2}\,\sum^{2N_e}_{l=1}\omega'_{l}\Theta(-\omega'_{l})=-\frac{N}{2}\,\sum^{N_e}_{l=1}(E_{0}-E_{l})\label{Gcl_3}.
\end{equation}
In a similar way, the third term on the right-hand side of Eq. (\ref{gpotential}) is rewritten as follows \cite{Aichhorn_Arrigoni_1,Senechal_arXiv}:
\begin{equation}
\sum_{\tilde {\bf k}} {\rm Tr} \ln{[ -{\bf G}_{\rm VCA}({\tilde {\bf k}}) ]}
=\sum_{\tilde{{\bf k}}}\sum^{2N_e}_{l=1}\omega_{l}(\tilde{{\bf k}})\Theta [ -\omega_{l}(\tilde{{\bf k}}) ]-R\label{Gvca_1},
\end{equation}
where $\omega_{l}({\tilde {\bf k}})$ is the pole of the VCA Green's function ${\bf G}_{\rm VCA}({\tilde {\bf k}})$ with ${\tilde {\bf k}}$ being the wave vector in the Brillouin zone of the reference system. The details of the numerical method to find $\omega_{l}({\tilde {\bf k}})$ will be given in the next paragraph. With the help of Eqs. (\ref{Gcl_2})--(\ref{Gvca_1}), we obtain the following expression for the grand potential per site:
\begin{eqnarray}
\!\!\!\!\!\!\nonumber \frac{\Omega}{N}\,&=&\,\frac{E_{0}}{2}-\frac{1}{2}\,\sum^{N_e}_{l=1}(E_{0}-E_{l})\\
&+&\frac{1}{N}\sum_{\tilde{{\bf k}}}\sum^{2N_e}_{l=1}\omega_{l}(\tilde{{\bf k}})\Theta [ -\omega_{l}(\tilde{{\bf k}}) ]-2\,(\mu-\mu').
\end{eqnarray}
Here, the summation $\frac{1}{N}\sum_{\tilde{\bf k}}$ is replaced by the integration $\frac{1}{(2 \pi)^2} \int^{\pi/2}_{-\pi/2} d \tilde{k}_{x} \int^{\pi}_{-\pi} d \tilde{k}_{y} $ in thermodynamic limit $N\rightarrow \infty$.

We finally present the numerical method to find the poles of the VCA Green's function ${\bf G}_{\rm VCA}(\tilde{{\bf k}})$. The VCA Green's function ${\bf G}_{\rm VCA}(\tilde{{\bf k}})$ is given by \cite{Aichhorn_Arrigoni_1,Senechal_arXiv}
\begin{eqnarray}
\nonumber {\bf G}_{\rm VCA}(\tilde{{\bf k}})&=&\frac{1}{{\bf G}_{0}(\tilde{\bf k})^{-1}-{\bf \Sigma}'}\\
\nonumber &=&\frac{1}{{\bf G}_{0}(\tilde{\bf k})^{-1}-({{\bf G}'_{0}}^{-1}-{{\bf G}'}^{-1})}\\
\nonumber &=&\frac{1}{(\omega-{\bf T}(\tilde{{\bf k}}))-(\omega-{\bf T'}-{{\bf G}'}^{-1})}\\
&=&\frac{1}{{\bf G'}^{-1}-{\bf V}(\tilde{{\bf k}})}\label{Gvca_2},
\end{eqnarray}
where the matrices ${\bf T}(\tilde{{\bf k}})$ and ${\bf T'}$ are
\begin{equation}
{\bf T}(\tilde{{\bf k}})=\left(
\begin{array}{cc}
{\bf A}(\tilde{{\bf k}}) & {\bf 0} \\
{\bf 0} & -{\bf A}(\tilde{{\bf k}}) \\
\end{array}
\right),\ \ 
{\bf T'}=\left(
\begin{array}{cc}
{\bf B} & {\bf C} \\
{\bf C} & {\bf D} \\
\end{array}
\right),
\end{equation}
with
\begin{equation}
{\bf A}(\tilde{{\bf k}})=\left(
\begin{array}{cccc}
-\mu & \epsilon(\tilde{{\bf k}}) & -V & 0 \\
\epsilon^{\ast}(\tilde{{\bf k}}) & -\mu & 0 & -V \\
-V & 0 & \epsilon_{f} \!-\! \mu & 0 \\
0 & -V & 0 & \epsilon_{f} \!-\! \mu \\
\end{array}
\right),
\end{equation}
\begin{equation}
\epsilon(\tilde{{\bf k}})=-t\bigl(1+e^{-i2\tilde{k}_{x}}+e^{-i(\tilde{k}_{x}-\tilde{k}_{y})}+e^{-i(\tilde{k}_{x}+\tilde{k}_{y})}\bigr),
\end{equation}
\begin{equation}
{\bf B}=\left(
\begin{array}{cccc}
-\mu' & -t & -V & 0 \\
-t & -\mu' & 0 & -V \\
-V & 0 & \epsilon_{f} \!-\! \mu' \!-\! h'_{\rm AF} & 0 \\
0 & -V & 0 & \epsilon_{f} \!-\! \mu' \!+\! h'_{\rm AF} \\
\end{array}
\right),
\end{equation}
\begin{equation}
{\bf C}=\left(
\begin{array}{cccc}
-h'_{cc} & 0 & h'_{cf} & 0 \\
0 & -h'_{cc} & 0 & h'_{cf} \\
h'_{cf} & 0 & h'_{ff} & 0 \\
0 & h'_{cf} & 0 & h'_{ff} \\
\end{array}
\right),
\end{equation}
and
\begin{equation}
{\bf D}=\left(
\begin{array}{cccc}
\mu' & t & V & 0 \\
t & \mu' & 0 & V \\
V & 0 & -\epsilon_{f} \!+\! \mu' \!-\! h'_{\rm AF} & 0 \\
0 & V & 0 & -\epsilon_{f} \!+\! \mu' \!+\! h'_{\rm AF} \\
\end{array}
\right).
\end{equation}
The matrix ${\bf V}(\tilde{{\bf k}})\,\,{\equiv}\,\,{\bf T}(\tilde{{\bf k}})-{\bf T'}$ denotes the intercluster hopping. By substituting Eq. (\ref{Gcl_1}) into Eq. (\ref{Gvca_2}), we obtain
\begin{equation}
{\bf G}_{\rm VCA}(\tilde{{\bf k}})={\bf Q}\,\frac{1}{{\bf g}^{-1}-{\bf Q}^{\dagger}\,{\bf V}(\tilde{{\bf k}})\,{\bf Q}}\,{\bf Q}^{\dagger}.
\end{equation}
This expression shows that the poles of the VCA Green's function are given as the eigenvalues of the matrix ${\bf L}(\tilde{{\bf k}})={\bm \Lambda}+{\bf Q}^{\dagger}\,{\bf V}(\tilde{{\bf k}})\,{\bf Q}$.


\end{document}